\documentclass[3p]{elsarticle}

\usepackage[numbers]{natbib}
\usepackage{amssymb}
\usepackage{amsmath}
\usepackage{hyperref}
\usepackage{threeparttable}
\usepackage{aas_macros}
\usepackage{xcolor}

\begin{document}

\begin{frontmatter}

\title{A highly efficient Voigt program for line profile computation}

\author[1]{Mofreh R. Zaghloul\corref{cor1}
  \fnref{fn1}}
  \ead{M.Zaghloul@uaeu.ac.ae}

\author[2]{Jacques Le Bourlot}
  \ead{Jacques.Lebourlot@obspm.fr}

\cortext[cor1]{Corresponding author}
\fntext[fn1]{The Fortran90 code can be requested from the first author or downloaded from https://github.com/mofrehzaghloul.}

\affiliation[1]{organization={United Arab Emirates University},
            addressline={Dept. of Physics, College of Sciences}, 
            city={AlAin},
            postcode={15551}, 
            state={Abu Dhabi},
            country={UAE}}

\affiliation[2]{organization={LERMA, Observatoire de Paris, PSL Research University, CNRS, Sorbonne Universités, 75014 Paris, France \\ and Université Paris-Cité, Paris, France}}

\begin{abstract}
Evaluation of the Voigt function, a convolution of a Lorentzian and a Gaussian profile, is essential in various fields such as spectroscopy, atmospheric science, and astrophysics. Efficient computation of the function is crucial, especially in applications where the function may be called for an enormous number of times.
In this paper, we present a highly efficient novel algorithm and its Fortran90 implementation for the practical evaluation of the Voigt function with accuracy in the order of $10^{-6}$. 
The algorithm uses improved fits based on Chebyshev subinterval polynomial approximation for functions in two variables.
The algorithm significantly outperforms widely-used competitive algorithms in the literature, in terms of computational speed, making it highly suitable for real-time applications and large-scale data processing tasks.
The substantial improvement in efficiency positions the present algorithm and computer code as a valuable tool in relevant scientific domains. The algorithm has been adopted and implemented in the Meudon PDR code at Paris Observatory and is recommended for similar applications and simulation packages. \nocite{*}
\end{abstract}

\begin{highlights}
\item Highly efficient algorithm for evaluating the Voigt function with an accuracy of \(10^{-6}\).
\item Chebyshev subinterval polynomial approximation enhances computational performance.
\item Significant speed improvement over existing algorithms, suitable for real-time applications.
\item Optimized for large-scale data processing. 
\item Adopted by the Meudon PDR code for its proven superior efficiency. 
\end{highlights}


\end{frontmatter}



\section{Introduction}

The Voigt function has broad applicability in many fields, including spectroscopy, astronomy, atmospheric science, and plasma physics (see \cite{Armstrong1967, Rothman2009}). The function, which is the convolution of the Lorentzian and Gaussian profiles, is written mathematically as:
\begin{equation}
V(x, y) = \frac{y}{\pi} \int_{-\infty}^{\infty} \frac{e^{-t^2}}{y^2 + (x - t)^2} \, dt, \quad y > 0
\label{Eq:01}
\end{equation}

The dimensionless variables $x$ and $y$ are defined in terms of the distance to line center $(\nu - \nu_0)$ and of the Lorentzian and Doppler half-widths $\gamma_L$ and $\gamma_D$:
\begin{equation}
x = \sqrt{\ln 2} \, \frac{\nu - \nu_0}{\gamma_D}, \quad y = \sqrt{\ln 2} \, \frac{\gamma_L}{\gamma_D}
\label{Eq:02}
\end{equation}

In these equations, $\nu_0$ is the frequency at the line center, and $\nu$ is the frequency at the point of calculation. It is also easy to recognize that the function is even in $x$.

Accurate and efficient evaluation of the Voigt function is essential for simulations and analyses in various fields. Many applications require evaluating the Voigt function over large datasets or a vast number of points making computational efficiency crucial for processing this volume within a reasonable timeframe. Moreover, in real-time data analysis or control systems, rapid evaluation of the Voigt function is necessary to deliver timely results. Efficient algorithms are also beneficial in environments with limited computational resources, such as embedded systems or mobile devices, where performance and energy consumption are critical factors.

In addition to computational efficiency, considerable accuracy in evaluating the Voigt function is essential for the stability of simulations and minimizing error propagation in large-scale analyses and applications. For many practical applications in the aforementioned fields, an accuracy on the order of $10^{-5}$ or $10^{-6}$ is useful. 

The most efficient algorithm and computer code known to the present authors that achieves this level of accuracy is the HUMLIK code developed by \cite{Wells1999}. This algorithm is compiled and adapted from various existing algorithms to produce a more efficient code for atmospheric line-by-line calculations. The HUMLIK code is adopted in many software packages, such as the Meudon PDR code \cite{2006ApJS..164..506L}, developed at the Paris Observatory. The Meudon PDR code is used to study various interstellar environments including diffuse clouds, dark clouds, and photon-dominated regions, providing valuable insights into the physics and chemistry of these regions. The HUMLIK code builds on, and improves the methodologies used in \cite{Humlicek1982} W4 and \cite{Humlicek1979} CPF12 algorithms, computing the Voigt function using real arithmetic to an accuracy on the order of $10^{-6}$ and with greater efficiency. As mentioned above, the development of highly efficient codes for evaluating the Voigt function is driven by the need to handle the increasing volume and complexity of data in modern scientific research. 

The objective of the present work is to address this need by developing and introducing a highly efficient algorithm and associated Fortran90 code for evaluating the Voigt function. The present algorithm presents a significant efficiency improvement compared to the HUMLIK code while maintaining accuracy on the order of $10^{-6}$. 

In Sect.~\ref{Algorithm} we provide a description of the proposed algorithm, highlighting the key new techniques that contribute to its superior efficiency. In Sect.~\ref{Implementation} we present a comprehensive evaluation of the algorithm's performance, including comparisons with Wells’ HUMLIK algorithm in terms of computational speed -- CPU time -- for different sets of data points. In Sect.~\ref{Application}, the application of the present algorithm in modeling Photon Dominated Regions (hereafter PDR) using the so-called "Meudon PDR  code" \footnote{Available at \href{https://pdr.obspm.fr}{https://pdr.obspm.fr}} is described and discussed. Finally, in Sect.~\ref{Conclusions}, we conclude by summarizing our results and discussing the potential applications of our work. 

\section{Algorithm}\label{Algorithm}

The Voigt function, which is the real part of the Faddeyva/Faddeeva function $ w(z=x+iy)$, can be asymptotically approximated by a continued fraction \cite{Faddeyeva1961, Abramowitz1964, Gautschi1970, Zaghloul2016, Zaghloul2017}:
\begin{equation}
w(z) \approx \frac{i}{\sqrt{\pi}} \, \frac{1}{z -} \, \frac{\frac{1}{2}}{z -} \, \frac{1}{z -} \frac{\frac{3}{2}}{z -} \, \frac{2}{z -} \dots
\label{Eq:3}
\end{equation}
This continued fraction needs to be truncated at some convergent for practical evaluation.

Using our recently published algorithm for calculating the Faddeyeva function to $14$ significant digits \cite{Zaghloul2024} as a reference, we specify the regions of applicability corresponding to the number of convergents retained (up to five) for an accuracy on the order of $10^{-6}$. However, since the CPU execution time increases with the number of convergents retained, we choose the minimum number necessary to achieve the specified accuracy wherever these regions overlap.

The first five rows in Table~\ref{table:1} summarize the proposed regions of applicability and the equivalent rational approximation up to five convergents of Laplace continued fractions. 

\begin{table}[]
\centering

\begin{threeparttable}
\caption{Regions of the proposed partitioning and the corresponding computational
method used, where $z=x+iy$.}

\label{table:1}
\begin{tabular}{ccc}
\hline
{\small Region } & {\small Borders } & {\small Method }\tabularnewline
\hline
{\small I } & {\small$(x^{2}+y^{2})\geq1.6 \times 10^{5}$ } & {\small 1 convergent }\tabularnewline
 &  & {\small$V(x,y)\approx\frac{y}{\sqrt\pi\,(x^{2}+y^{2})}$ }\tabularnewline
{\small 
II } & {\small $1.6 \times 10^{5}>(x^{2}+y^{2})\geq510.0$ } & {\small 2 convergents }\tabularnewline
 &  & {\small$V(x,y)\approx\frac{y\,(0.5+x^{2}+y^{2})}{\sqrt\pi\,((x^{2}+y^{2})^{2}+y^{2}-x^{2}+0.25)}$ }\tabularnewline
{\small 
III } & {\small$510.0>(x^{2}+y^{2})\geq110.0$ } & {\small 3 convergents \tnote{1} }\tabularnewline
 &  & {\small$V(x,y)\approx\Re\left(\frac{i\,(z^{2}-1)}{z\,\sqrt\pi\,(z^{2}-1.5)}\right)$ }\tabularnewline
{\small 
IV } & {\small$110.0>(x^{2}+y^{2})\geq38.0$ } & {\small 4 convergents \tnote{1,2} }\tabularnewline
 & {\small\& $y>10^{-8}$ } & {\small$V(x,y)\approx\Re\left(\frac{i\,z\,(z^{2}-2.5)}{\sqrt\pi\,(z^{2}\,(z^{2}-3)+0.75)}\right)$ }\tabularnewline
{\small 
V } & {\small$38.0>(x^{2}+y^{2})\geq25.0$ } & {\small 5 convergents \tnote{1,2} }\tabularnewline
 & {\small\& $y>10^{-3}$ } & {\small$V(x,y)\approx\Re\left(\frac{i\,(z^{2}\,(z^{2}-4.5)+2.0)}{z\,\sqrt\pi\,(z^{2}\,(z^{2}-5.0)+3.75)}\right)$ }\tabularnewline
{\small 
VI } & {\small$25>(x^{2}+y^{2})\geq7.3916$ } & {\small Chebyshev subinterval }\tabularnewline
 & {\small\& $y>2 \times 10^{-2}$ } & {\small bivariate polynomial }\tabularnewline
 &  & {\small approximations\tnote{2} }\tabularnewline
{\small 
VII } & {\small$7.3916>|z|^{2}$ } & {\small Chebyshev subinterval }\tabularnewline
 &  & {\small bivariate polynomial }\tabularnewline
 &  & {\small approximations }\tabularnewline
\hline
\end{tabular}

\begin{tablenotes}
    \item[1]{Explicit expressions in
real arithmetic are used for the real parts of the complex expressions
for $w(z)$.}
    \item[2]{For the corresponding missing
regions with small values of $y$, a few terms of the Taylor series
of the Dawson integral are used to calculate the function to the sought
accuracy.}
\end{tablenotes}

\end{threeparttable}

\end{table}

As shown in the table, the real part of these convergents can be used to approximate the Voigt function to an accuracy on the order of $10^{-6}$ for $x^2 + y^2 \geq 25$, except for the narrow strips defined by $110 > (x^2 + y^2) \geq 38$ and $y \leq 10^{-8}$, and $38 > (x^2 + y^2) \geq 25$ and $y \leq 10^{-3}$. In these narrow strips, a few terms of the truncated Taylor series of the Dawson integral, Daw(z) (expansion near $z_0 = x$) \citep{Zaghloul2019} can be used, where:
\begin{equation}
 w\left(x + i \delta y\right) \approx e^{-\left(x + i \delta y\right)^2} + \frac{2i}{\sqrt{\pi}} \left[d_0 + d_1 \left(i \delta y\right) + d_2 \left(i \delta y\right)^2 + \cdots\right]
\label{Eq:04}
\end{equation}
The expansion coefficients can be calculated recursively \cite{Armstrong1967, Zaghloul2019}:
\begin{flalign}
d_{0} & =\mathrm{Daw}\left(x\right)\,;\quad d_{1}=1-2\,d_{0}\nonumber\\
d_{n+1} & =\frac{2}{n+1}\,\left(x\,d_{n}+d_{n-1}\right),\quad n=1,2,\dots
\label{Eq:05}
\end{flalign}

Different simple approximations for $\mathrm{Daw}(x)$ are used depending on the $x$-values. For example, for $x \geq 5.0$, a sufficient number of terms from the following polynomial asymptotic series can be used to satisfy the sought accuracy:
\begin{equation}
\mathrm{Daw}(x) \approx \frac{1}{2x} \sum_{j=0}^{\infty} \frac{(2j - 1)!!}{(2x^2)^j} \approx \frac{1}{2x} + \frac{1}{4x^2} + \frac{3}{8x^3} + \frac{15}{16x^4} + \dots \label{Eq:06}
\end{equation}
where "!!" refers to the double factorial defined by $n!! = n \, (n - 2) \, (n - 4) \, (n - 6) \dots$ with $0!! = 1$ and $(-1)!! = 1$.

For the region $x^2 + y^2 < 25$, and except for the narrow strip defined by $2.71875 \leq x < 5.0$ and $y < 0.01831$, a two-variable Chebyshev subinterval polynomial approximation has been used to derive a set of bivariate polynomials to approximate the Voigt function to the required accuracy.

The classic technique of Chebyshev subinterval polynomial approximation is straightforwardly extended to the case of functions of two variables. For a given continuous function of two variables $f(y_1, y_2)$ with $y_1, y_2 \in [-1, 1]$, one can have a Chebyshev series \cite{Reutskiy2006, Scheiber2015, Malachivskyy2021, Malachivskyy2022, Malachivskyy2023}:
\begin{equation}
f(y_1, y_2) = \sum_{n=0}^{\infty} \sum_{m=0}^{\infty} \gamma_{n,m} \, T_n(y_1) \, T_m(y_2) \label{Eq:07}
\end{equation}
Here, $\gamma_{n,m}$ represent the coefficients of the Chebyshev series, and $T_k(y)$ is the Chebyshev polynomial of the first kind and order $k$.

For functions in two variables, $f(x_1, x_2)$, defined in semi-infinite domains $[0, \infty]$ or infinite domains $[-\infty, \infty]$, it is likely to perform linear and nonlinear space transformations. For example, when $x_1, x_2 \in [0, \infty]$, a nonlinear mapping may be advised to map the variables first to $t_1$ and $t_2$ on the finite domain $[0, 1]$, where:
\begin{equation}
t_1 = \frac{c_1}{x_1 + c_1}, \quad t_2 = \frac{c_2}{x_2 + c_2} \label{Eq:08}
\end{equation}
with $c_1$ and $c_2$ as constants. The constants $c_1$ and $c_2$ were both chosen to be 1.8125 based on numerical experiments. However, in general for the application of the technique, they do not have to be equal.

On the other hand, when the variables $t_1$ and $t_2$ belong to the finite interval $[a_1, b_1]$ and $[a_2, b_2]$, respectively, linear space coordinate transformation may be applied to map them into the range $[-1, 1]$:
\begin{equation}
y_1 = \frac{2 \, t_1 \, (x_1) - (b_1 + a_1)}{b_1 - a_1}, \quad y_2 = \frac{2 \, t_2 \, (x_2) - (b_2 + a_2)}{b_2 - a_2} \label{Eq:09}
\end{equation}
Depending on the arithmetic under consideration and the desired level of accuracy, the domains of $t_1$ and $t_2$ are partitioned into a finite number of sub-regions (not necessarily of equal sizes). In each sub-region, truncated series of bivariate Chebyshev polynomials $P(t_1, t_2)$ or $P(y_1, y_2)$, are obtained to approximate the function.

Making use of the orthogonality of the Chebyshev polynomials, $T_k(x) = \cos(k \cos^{-1}(x))$, where $x = \cos \theta$, and following \cite{Scheiber2015}, the coefficients of the polynomials $P(y_1, y_2)$ are given by:
\begin{equation}
\gamma_{0,0} = \frac{1}{\pi^2} \int_{-1}^{1} \int_{-1}^{1} \frac{f(y_1, y_2)}{\sqrt{1 - y_1^2} \sqrt{1 - y_2^2}} \, dy_1 \, dy_2, \label{Eq:10}
\end{equation}
\begin{equation}
\gamma_{n,0} = \frac{2}{\pi^2} \int_{-1}^{1} \int_{-1}^{1} \frac{f(y_1, y_2) \, T_n(y_1)}{\sqrt{1 - y_1^2} \, \sqrt{1 - y_2^2}} \, dy_1 \, dy_2, \label{Eq:11}
\end{equation}
\begin{equation}
\gamma_{0,m} = \frac{2}{\pi^2} \int_{-1}^{1} \, \int_{-1}^{1} \frac{f(y_1, y_2) \, T_m(y_2)}{\sqrt{1 - y_1^2} \, \sqrt{1 - y_2^2}} \, dy_1 \, dy_2, \label{Eq:12}
\end{equation}
\begin{equation}
\gamma_{n,m} = \frac{4}{\pi^2} \, \int_{-1}^{1} \, \int_{-1}^{1} \, \frac{f(y_1, y_2) \, T_n(y_1) T_m(y_2)}{\sqrt{1 - y_1^2} \, \sqrt{1 - y_2^2}} \, dy_1 \, dy_2 \label{Eq:13}
\end{equation}
Efficient evaluation of the coefficients in Eq.\ref{Eq:10} -- \ref{Eq:13} is possible using discrete Fourier transform where:
\begin{align}
\gamma_{n,m} = \frac{\alpha}{(N+1)(M+1)} 
\sum_{i=0}^{N} \sum_{j=0}^{M} f(y_1, y_2) \cos\left( \frac{n\,(2i+1)\,\pi}{2\,(N+1)} \right) \cos\left( \frac{m\,(2j+1)\,\pi}{2\,(M+1)} \right). \label{14}
\end{align}

where $\alpha = 4$ for $n \neq 0$ and $m \neq 0$, $\alpha = 2$ for $n = 0$ and $m \neq 0$ or $n \neq 0$ and $m = 0$, and $\alpha = 1$ for $n = 0$ and $m = 0$.

For practical evaluation of the series in Eq.~\ref{Eq:07}, one may terminate the sums to the powers $N$ and $M$ for $y_1$ and $y_2$, respectively, i.e.
\begin{equation}
f(y_1(x_1), y_2(x_2)) \approx \sum_{n=0}^{N} \, \sum_{m=0}^{M} \, \gamma_{n,m} \, T_n(y_1) \, T_m(y_2)
\label{Eq:15}
\end{equation}
Values of the degrees of the bivariate polynomials, denoted by $N$ and $M$, are generally chosen to satisfy a targeted accuracy of the approximation. They depend not only on the characteristics of the function $f(x_1, x_2)$ but also on the sub-region, in addition to the desired accuracy.
Components and regions of the proposed partitioning, as well as the computational methods employed in the current algorithm are shown in Table~\ref{table:1}.

There remains a small strip to close the domain of computation that is for $2.71875 \leq |x| < 5$, and $y \leq 0.0180808080808088$. A polynomial of degree $12$ has been derived to approximate $\mathrm{Daw}(x)$ with accuracy better than $10^{-8}$ using Chebyshev polynomial approximation. The expression is also accurate up to $x = 7.25$. 

\section{Implementation, Accuracy Check and Efficiency Benchmark}\label{Implementation}

The above-described algorithm has been implemented as a Fortran90 subroutine that receives as input a vector of real variables $x$ and a real scalar $y$ and returns as output the Voigt function $V(x,y)$. A recently published algorithm for calculating the Faddeyeva function up to $14$ significant digits using double precision arithmetic \citep{Zaghloul2024} is used as a reference to check the accuracy of the present algorithm and Fortran90 code.

\begin{figure}[h]
   \centering 
   \includegraphics[width=0.6\textwidth]{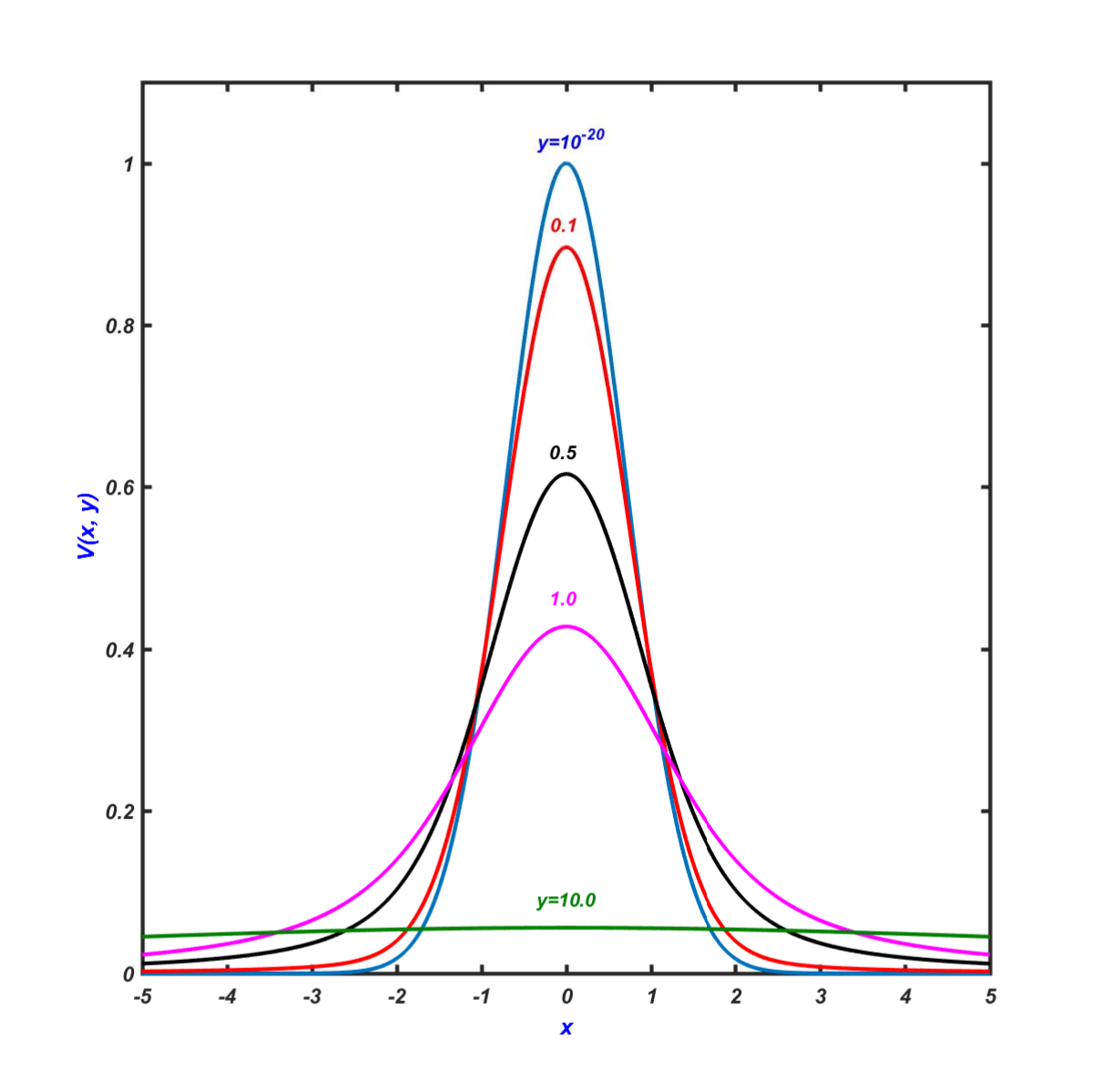} 
   \caption{Voigt function for different values of the parameter $y$ calculated using the present algorithm.}
   \label{Voigt Profile}
\end{figure}

Fig.~\ref{Voigt Profile} shows sample calculations of the Voigt function using the present algorithm for values of $x \in [-5, 5]$ and different values of the parameter $y$.

To perform a thorough accuracy check and speed benchmark, an array of $40\,000$ linearly spaced values between $-x_{\mathrm{max}}$ and $x_{\mathrm{max}}$ is used. The computations are repeated $45$ times, each with a different value of $y$, starting from $y_{\mathrm{min}}$ and multiplied by $\sqrt{10}$ each time. Table~\ref{tab:comparison} shows a summary of a set of numerical experiments and tests showing a comparison between the present algorithm and Fortran90 computer code and the HUMLIK algorithm and FORTRAN77 computer code. As can be seen from the results in the table, both algorithms and Fortran codes secure accuracy on the order of $10^{-6}$ for the tested data sets. However, the Fortran90 implementation of the present algorithm is considerably faster than the HUMLIK algorithm. Efficiency improvements by factors greater than $2$ and up to an order of magnitude are obtained depending on the values of $x_{\mathrm{max}}$ and $y_{\mathrm{min}}$.

\begin{table}

\caption{Comparison between the present algorithm and computer code ($P$) and the Humlíček algorithm and computer code ($H$). Computations were performed using the GNU \textit{gfortran} compiler 8.1.0 on an Intel(R) Core(TM) i7-6600U CPU @ 2.60GHz, 2.81 GHz.} \protect\label{tab:comparison}

\centering%
\begin{tabular}{c c|c c|c c c}
\hline\hline
\multicolumn{2}{c}{Case} & \multicolumn{2}{c}{$\left|\mathrm{Relative \  Error}\right|_{\mathrm{max}}$} & \multicolumn{3}{c}{CPU Run Time ($\mathrm{s}$)}\tabularnewline
\hline
$x_{max}$ & $y_{min}$ & $H$ & $P$ & $H$ & $P$ & $H/P$\tabularnewline
\hline
$20$ & $10^{-10}$ & $7.9\,10^{-6}$ & $9.1\,10^{-6}$ & $4.02$ & $2.00$ & $2.01$\tabularnewline
$100$ & $10^{-10}$ & $8.9\,10^{-6}$ & $9.7\,10^{-6}$ & $6.06$ & $0.81$ & $7.48$\tabularnewline
$200$ & $10^{-10}$ & $9.0\,10^{-6}$ & $9.7\,10^{-6}$ & $6.23$ & $0.70$ & $8.90$\tabularnewline
$20$ & $10^{-20}$ & $7.9\,10^{-6}$ & $9.1\,10^{-6}$ & $10.78$ & $4.55$ & $2.37$\tabularnewline
$100$ & $10^{-20}$ & $8.9\,10^{-6}$ & $9.7\,10^{-6}$ & $20.42$ & $1.42$ & $14.38$\tabularnewline
$200$ & $10^{-20}$ & $9.0\,10^{-6}$ & $9.7\,10^{-6}$ & $21.05$ & $1.03$ & $20.44$\tabularnewline
\hline
\end{tabular}

\end{table}

\section{Application to PDRs}\label{Application}

Radiative transfer in atomic and molecular lines plays a fundamental role in the physics of PDRs (Photon Dominated Regions, of the Interstellar Medium), where atomic hydrogen is converted to molecular hydrogen. An accurate method to deal with the problem of anisotropic scattering in line transfer is described in \cite{2007A&A...467....1G}. However, to be able to compute wavelength dependent absorption and emission coefficients, a preliminary necessary step is to account for lines profiles. $\mathrm{H}_2$ lines develop very wide lines where Lorentzian wings may extend over several $\mathrm{nm}$ in the far ultraviolet range. Thus, using a efficient Voigt profile routine is mandatory.

$\mathrm{H}_2$ fundamental electronic states $X\ ^1\Sigma^+_g$ has $302$ ro-vibrational levels, which leads to about $36400$ transitions up to one of the six lowest electronic excited states $B\ ^1\Sigma^+_u$, $C^+\ ^1\Pi_u$, $C^-\ ^1\Pi_u$, $B'\ ^1\Sigma^+_u$, $D^+\ ^1\Pi_u$ and $D^-\ ^1\Pi_u$. Radiative pumping to one of the $2135$ accessible levels is followed by radiative decay whith a rich fluorescence spectrum. Only $10\%$ of these decays lead to dissociation of the molecule, which explains why shielding of the radiation field by $\mathrm{H}_2$ lines dominates the radiative energy density in the outer layers of a PDR.

   \begin{figure}
   \centering
   \includegraphics[width=0.8\columnwidth]{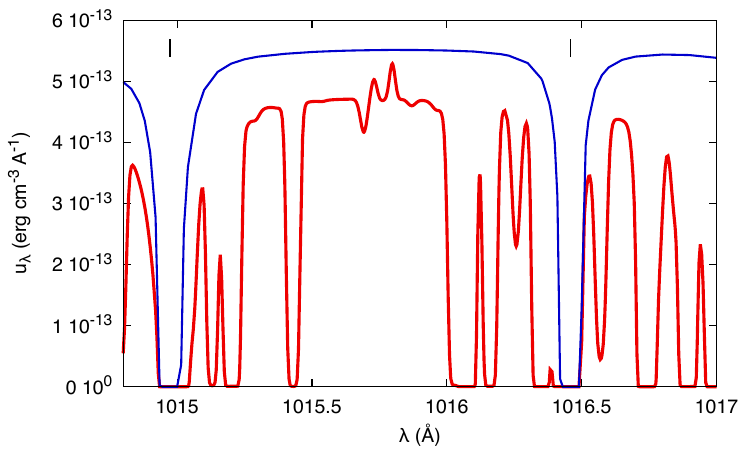}
      \caption{Radiative energy density at $A_{\mathrm{V}} = 0.1$, using only $4$ levels of $\mathrm{H}_2$ (blue) or the full set of possible transitions (red). The short ticks show the positions of absorptions from level $(v=0,J=2)$.}
         \label{UV_field}
   \end{figure}

   \begin{figure}
   \centering
   \includegraphics[width=0.8\columnwidth]{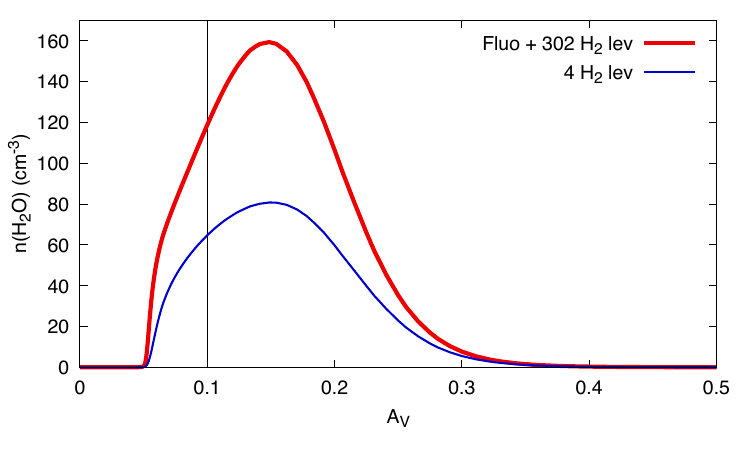}
      \caption{Abundance of $\mathrm{H_2O}$ molecules in case 1 (blue) and case 2 (red). The vertical bar corresponds to the position where the radiation field of Fig.\ref{UV_field} is shown.}
         \label{H2O_ab}
   \end{figure}

As only a handful of ro-vibrational levels are significantly populated, one could be tempted to include only a few lines into the radiative transfer computation. This may lead to significant errors as we show here. We model a strong PDR, typical of conditions close to the Orion bar using the Meudon PDR code (see \cite{2006ApJS..164..506L} for a description of the code, and \cite{2024A&A...689L...4G} for the model used). Line transfer is performed using either only $\mathrm{H}_2$ $4$ lowest rotational levels (case 1) or the full set of $302$ levels (case 2). Fig.~\ref{UV_field} shows a small portion of the radiative energy density $u_{\lambda}$ at a depth of $A_{\mathrm{V}} = 0.1 \, \mathrm{mag}$ into the cloud. Using more levels reduces significantly the energy density. We observe three effects:

   \begin{itemize}
      \item The lowest levels lead to strong absorption where no radiation is left at line centre and whose wings extend far.
      \item Excited levels have multiple weak lines that overlap. This leads to an overall reduction of the energy density over the whole UV range.
      \item Fluorescence adds weak emission features due to photons reemitted towards the deeper part of the cloud. This increases slightly the energy density compared to what is found with pure absorption.
   \end{itemize}

The physical consequences are illustrated on Fig.~\ref{H2O_ab}, which shows the abundance of the $\mathrm{H_2O}$ molecule just beyond the outer layers of the PDR, in a region where the gas kinetic temperature (about $1000\ \mathrm{K}$) is still high enough to lead to multiple lines emission. Failing to include all levels of $\mathrm{H}_2$ leads to a computed abundance of $\mathrm{H_2O}$ which is divided by a factor of $2$. This is because overestimating the radiation field energy density leads to a proportional overestimate of $\mathrm{H_2O}$ photodissociation rate, which is computed by integrating the wavelength dependent cross-section over $u_{\lambda}$.

Accounting for these physical processes would not be possible without an accurate and fast routine for the computation of the Voigt profile.

\section{Conclusions}\label{Conclusions}

A highly efficient algorithm is introduced, along with its corresponding Fortran90 implementation, for practical evaluation of the Voigt function with an accuracy of up to $10^{-6}$. The present algorithm significantly enhances computational efficiency, outperforming the widely-used HUMLIK algorithm by Wells.
Specifically, the present algorithm consistently achieves speed-ups by factors greater than $2$ and up to a factor of $20$ compared to the HUMLIK algorithm, depending on the specific values of the input parameters $x$ and $y$. This remarkable efficiency makes it particularly suitable for applications requiring rapid and repeated evaluations of the Voigt function. 

The Fortran code implementation of this algorithm is optimized for performance, ensuring seamless integration into existing computational packages requiring efficient computation of the Voigt function. Researchers interested in this implementation can request the Fortran code for their use.

Moreover, the techniques introduced in this algorithm can be used to calculate the imaginary part of the Faddeyeva function, further broadening applicability, which could be considered in future work.

\section*{Acknowledgements}
   This work is supported by the UAE University UPAR research grant number 2278, 2023. JLB acknowledge support from the Programme National ``Physique et Chimie du Milieu Interstellaire" (PCMI) of the CNRS/INSU with INC/INP co-funded by CEA and CNES.

\bibliographystyle{elsarticle-num}
\bibliography{Voigt_JQSRT}

\end{document}